\documentclass[conference]{IEEEtran}
\IEEEoverridecommandlockouts
\IEEEoverridecommandlockouts
\IEEEpubid{\makebox[\columnwidth]{ 979-8-3503-6583-
2/25/\$31.00~\copyright~2025~European Union \hfill}
\hspace{\columnsep}\makebox[\columnwidth]{ }}

\usepackage{cite}
\usepackage{amsmath,amssymb,amsfonts}
\usepackage{algorithmic}
\usepackage{graphicx}
\usepackage{textcomp}
\usepackage{xcolor}
\usepackage{subfig}

\def\BibTeX{{\rm B\kern-.05em{\sc i\kern-.025em b}\kern-.08em
    T\kern-.1667em\lower.7ex\hbox{E}\kern-.125emX}}

\begin{document}

\title{Single Antenna Terahertz Sensing using Preconfigured Metasurfaces
\thanks{
This work is funded by the German Research Foundation (“Deutsche Forschungsgemeinschaft”) (DFG) under Project–ID 287022738 TRR 196 for Project S03.} }

\author{\IEEEauthorblockN{Furkan H. Ilgac}
\IEEEauthorblockA{\textit{Ruhr University Bochum} \\
Bochum, Germany \\
Furkan.Ilgac@ruhr-uni-bochum.de}
\and
\IEEEauthorblockN{Aydin Sezgin}
\IEEEauthorblockA{\textit{Ruhr University Bochum} \\
Bochum, Germany \\
Aydin.Sezgin@ruhr-uni-bochum.de}}

\maketitle

\begin{abstract}
The development of mobile terahertz (THz) sensing and localization with minimal infrastructure has garnered significant attention due to its substantial practical implications. Single-antenna radar systems are a favored choice for mobile platforms, as they offer notable advantages in terms of cost, weight, and simplicity. However, these systems face a critical limitation: the inability to extract angular information using a single antenna, which consequently prevents the achievement of complete localization. This paper proposes an angular estimation method for a single-antenna radar augmented with a pair of preconfigured metasurfaces. The metasurface pair is used for creating an interference pattern in the scene, which depends on the target angles and operating frequency. Moreover, the beam squint effects caused by the wide frequency range in the THz band provides suitable conditions for using sparse reconstruction techniques to obtain angular estimates. We utilize these properties to perform angular estimation with a single antenna. The simulation results show that with this method it is  possible to perform fast and  accurate multi-target estimation for a broad operating range.
%\textit{This scienfitic paper is an outcome of the MARIE project( for more information, see trrmarie.de)[S03]} 
\end{abstract}


\section{Introduction}
    Advancements in electronics are enabling the utilization of new electromagnetic frequency bands. Among these, the terahertz band stands out as particularly significant due to its potential for high communication speeds, sub-millimeter localization accuracy, and material characterization capabilities. These features make it especially promising for mobile platforms involved in critical applications such as disaster management and emergency rescue. For instance, a mobile platform (e.g. drone) leveraging terahertz technology could access areas inaccessible to humans, achieve detailed self-localization and sensing of the environment, and even material characterization of the objects in the scene. Naturally, the primary requirement for such an autonomous drone is the precise localization. However, the highly directive propagation characteristics of terahertz waves pose challenges, particularly in non-line-of-sight (NLOS) or partially obstructed scenarios. To address these challenges, reconfigurable intelligent surfaces (RIS) have emerged as a leading solution.  

A reconfigurable intelligent surface (RIS) is a nearly-passive metasurface composed of densely arranged  antennas. By controlling the impedance across its surface, RIS enables controlled  management of the electromagnetic waves in the environment, thereby shaping the spatial spectrum. Initially introduced as a supportive tool for wireless communications \cite{kevin,yasemin}, RIS evolved to include applications in sensing and material characterization, significantly expanding their range of use cases. Currently, RIS-aided radar sensing applications is an active research field. The  capabilities obtained by RIS has been used for radar imaging \cite{ZhuHan, ZhuHan2,ilgac2, codedAperture}  enabling NLOS sensing\cite{ilgac,siri}, improving detection \cite{buzzi} and accuracy \cite{strob,henkAngleImp}. While RIS inherently operates without any RF chains, resulting in negligible power consumption, its is not entirely zero, still necessitating certain infrastructural support.

To reduce infrastructure requirements under harsh sensing conditions, the integration of novel passive and semi-passive RF structures has recently garnered significant interest. In this regard, RF tags has been proposed to provide reference anchors \cite{jesus,absi} improve trajectory accuracy \cite{marieSAR} and convey spatial information \cite{trafficSign}. Similarly, passive preconfigured metasurfaces have been suggested to enhance radar coverage in blocked areas \cite{dortmund,Ehsan}. These surfaces are tailored for a single configuration and primarily function as non-specular reflectors, offering drastic SNR improvements under NLOS conditions. 

Since high-frequency RF tags and metasurfaces are still in early development, most research focuses on device technology while being employed in relatively simple tasks, mainly as signal reflectors. Therefore, extending the capabilities of these devices by employing them directly in a signal processing problem remains relatively unexplored. In this study, we employ a simple single-antenna radar system in conjunction with two preconfigured surfaces to achieve both angle and range estimation. Although a single-antenna radar can only provide  range information, we leverage the high spectral bandwidth of the terahertz band to modify the interference pattern created by the preconfigured surfaces. By examining the change of the target echo under changing interference pattern, we successfully extract angle information in addition to range estimation, enabling full localization of the objects.

%\textcolor{red}{SISO radar ile neler yapilmis, nasi yapilmis}
% https://arxiv.org/pdf/2407.11726
% https://arxiv.org/pdf/2401.06544
% https://arxiv.org/html/2408.09883v1
%

\section{System Model}
\begin{figure}[h]
    \centering
    \includegraphics[width=0.70\linewidth]{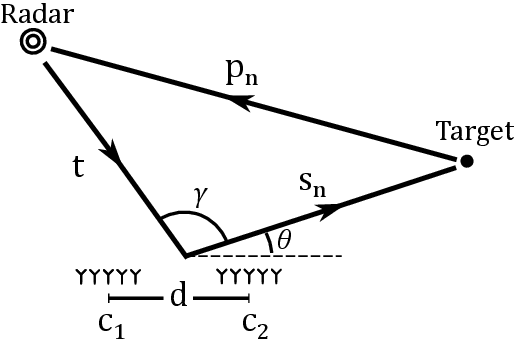}
    \caption{The proposed system setup.}
    \label{fig:system}
\end{figure}

    An illustration of the system geometry is given in Fig. \ref{fig:system}, which demonstrates a monostatic single antenna radar setup equipped by two preconfigured surfaces denoted as $c_1$ and $c_2$, placed along the same axis, separated distance $d$ apart. The radar performs sensing by usual transmissions and echo collections. With a single antenna, it can perform range estimations and calculate the distance $p_n$. However, to fully localize a target, the angular information of the targets are needed. To this end, the secondary path over the preconfigured surfaces is utilized. In the secondary path, the transmitted signals first reach the surfaces, where they are reflected, then travel to the target, and finally return to the radar, following the $t \xrightarrow{} s_n \xrightarrow{} p_n $ path.

Now, suppose that the radar employs frequency modulated continuous wave (FMCW) signals with unit-amplitude for sensing the environment, which is given as
\begin{equation}
y_{\text{RF}}(\tau) = \exp \left( j 2\pi f_c \tau + j \pi \frac{\beta}{T} \tau^2 \right),
\end{equation}
where  $f_c$ is the carrier frequency, $\frac{\beta}{T}$ is the chirp rate, which is the ratio of the  bandwidth $\beta$ to signal period $T$, and $\tau$ is fast-time. The signal reflected from a point-target located at a distance  $R_0$ from the radar can be modeled  as 
\begin{equation}
y_{\text{RF}}(\tau-\tau_o) =  \rho_{R_0}
y_{\text{RF}}(\tau) e^{ -j 2\pi f_c \tau_o - j\frac{2 \pi \beta }{T} \tau_o \tau + j\frac{\pi \beta}{T} \tau_o^2 }.
\end{equation}
The time shift $\tau_0 $ depends on the range $R_0$ and the speed of light $c$ following the relation $\tau_0 = R_0/c$. Following the down conversion, dechirping, and post-processing operations, the baseband equivalent of the signal model can be written as
\begin{equation}
y(R_0) =  \rho_{R_0}
\exp \left( -j k R_0 \right),
\end{equation}
where $k =  (2\pi f_c +2 \pi \gamma \tau)/c $ is the wavenumber, changing with the fast-time index \cite{richards}, and $\rho_{R_0} = \sigma/(4k^2R_0^2)$ denotes the total channel coefficient, modeled as the product of the radar cross section (RCS) $\sigma$, and the free-space path loss.

\begin{figure}[h] 
    \centering
    \includegraphics[width=0.80\linewidth]{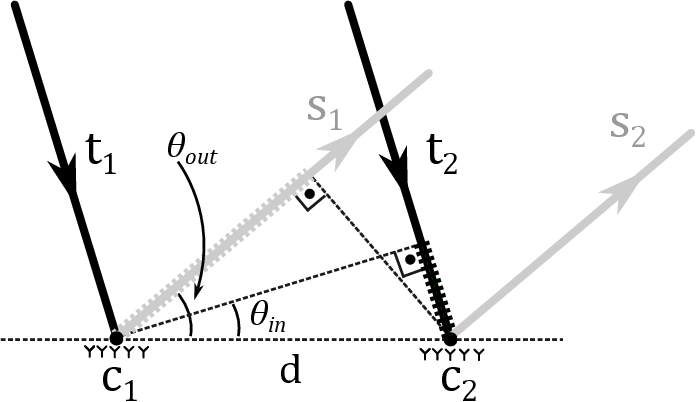}
    \caption{A close-up view of the setup in the far field showing the geometrical relations between the path lengths and angles. }
    \label{fig:system_closeup}
\end{figure}

Assuming that both the target and the radar are positioned in the far-field region of the preconfigured surface pair, a close-up view of the problem geometry is shown in Fig. \ref{fig:system_closeup}. Due to far-field assumption, the paths arriving and reflecting from the surfaces are assumed to be parallel, and the differences in these path lengths are denoted with the dashed segments in the figure. In this case, the differences among the path lengths depends on the angle of arrival $\theta_{\text{in}}$, and the angle of departure $\theta_{\text{out}}$ with the following relations
\begin{equation}
\begin{split} \label{eq:pathdiff}
    t_2 &= t_1 + d \cos{\theta_{\text{in}}}\\
    s_2 &= s_1 - d \cos{\theta_{\text{out}}}.
\end{split}
\end{equation}
Assuming the channels from the radar to surfaces and from surfaces to target are modeled with uniform linear array vectors denoted as $\textbf{a}_1$ and $\textbf{a}_2$, for a single point-target, the signal arriving to the receiver over the second path can be written as
\begin{equation}
\begin{split} 
    y_\theta = & \rho  [  \textbf{a}_1(\theta)^H
            \rm{\Psi}_1 
            \textbf{a}_1(\theta_{\text{in}})
            e^{-jk(t_1+s_{1}+p)}\\                    
    &+     \textbf{a}_2(\theta)^H
        \rm{\Psi}_2 
        \textbf{a}_2(\theta_{\text{in}})
        e^{-jk(t_2+s_{2}+p)} ] +n,
\end{split}
\end{equation}
where $n$ is the noise term, and $\rm{\Psi}_1$, $\rm{\Psi}_2$ denote the phase responses of the surfaces. Assuming both surfaces consist of same number of elements, the equation simplifies to
\begin{equation} \label{eq:compact_y}
\begin{split} 
    y_\theta = \rho
    \textbf{a}(\theta)^H
            \left(
            \textbf{a}_{\Psi_1}  
            +
        \textbf{a}_{\Psi_2}
        e^{-jk \Delta_{\theta}}
        \right) 
        e^{-jk(t_1+s_{1}+p)} + n,
\end{split}
\end{equation}
where the terms inside the parenthesis are written in concise form such that, $\textbf{a}_{\Psi} = \rm{\Psi}\textbf{a}(\theta_{\text{in}})$, and $ \Delta_{\theta} = d[\cos{\theta} -\cos{\theta_{\text{out}}}]$. Lastly, $\rho$ denotes the channel coefficient which includes radar cross section and atmospheric loss.

Extending this formulation to the multi-target case, the received signal becomes
\begin{equation} \label{eq:multi_y}
\begin{split} 
    y = \sum_m^M 
    \rho_m  
    \textbf{a}_{\theta_m}^H
            \left(
            \textbf{a}_{\Psi_1}  
            +
        \textbf{a}_{\Psi_2}
        e^{-jk \Delta_{\theta_m}}
        \right) 
        e^{-jk(t_1+s_{1,m}+p_m)} + n.
\end{split}
\end{equation}

Examining (\ref{eq:compact_y}), we observe that the amplitude of received signal depends on the channel coefficients, the beam patterns of the preconfigured surfaces, the instantaneous frequency, and most importantly the target angle. Assuming the target is stationary across one fast-time interval, by observing the change in the received signal amplitude as the  frequency parameter $k$ changes, we can deduce the target angle. To this end, we will present a brute force search, and a sparse reconstruction method.

\subsection{Brute Force Search}
The proposed algorithm is designed to extract the angular information following the range processing stage. Therefore, the position of the radar, the number of targets in the scene, their range parameters $p_m$ and the RCSs $\sigma_m$ assumed to be known. Under these assumptions, the problem reduces to estimating the optimal angular bins given a received signal for across different frequencies. To mathematically formulate the problem, all observations $y_\theta$ within a fast-time interval are concatenated into a vector $\textbf{y}$. A corresponding vector for a candidate angle $\theta$ is constructed based on the following model for its elements:
 \begin{equation} \label{eq:v}
    [v_\theta]_k = \rho_\theta \textbf{a}(\theta)^H
            \left(
            \textbf{a}_{\Psi_1}  
            +
        \textbf{a}_{\Psi_2}
        e^{-jk \Delta_{\theta}}
        \right) 
        e^{-jk(t_1+s_{1}+p)}.
\end{equation}
For the single-target case, the problem can now be formulated as follows: 
 \begin{equation} \label{eq:opt}
 \begin{aligned}
      \underset{\theta }{\min} \quad &\| 
      \textbf{y}
      -     \textbf{v}_\theta
      ||_2,\\
          \text{s.t} \quad & s_{1} = t\cos\gamma + \sqrt{t^2 - p^2\sin^2\gamma}\\
        &\rho_\theta = \sigma/(64k^6t^2s_1^2p^2).
 \end{aligned}
 \end{equation}
 where the constraint on $s_1$ is dictated by geometric relationships, derived using the cosine theorem. For the single-target problem, the solution can be found via a search over the target angle $\theta$. However in multi-target case, since the eq. (\ref{eq:multi_y}) is highly non-convex, the search must be performed for all combinations of angle-range $(\theta_m,p_m)$ tuples. Consequently, the problem complexity grows exponentially with the number of targets in the scene.

\subsection{Sparse Reconstruction} \label{chp:cs}
Since for most practical cases, the targets are sparsely distributed across the scene, we can utilize this a priori information to obtain super-resolution performance and accelerate the processing speed. To this end, for a given range, we construct a sensing matrix $\textbf{V}$ by combining model vectors from the eq. (\ref{eq:v}) for all candidate angles. Arranging a linear equation system for the received signal with this sensing matrix yields
\begin{equation}
\begin{bmatrix}
y_{k_1} \\
\vdots \\
y_{k_K}
\end{bmatrix}
=
\begin{bmatrix}
\vdots & \cdots & \vdots \\
\hat{v}_{1} & \cdots & \hat{v}_{\Theta} \\
\vdots & \cdots & \vdots \\
\end{bmatrix}
\begin{bmatrix}
\sigma_1 \\
 \\
\vdots \\
\sigma_\Theta
\end{bmatrix},
\end{equation}
where $\hat{v}_\theta$ vectors are in the same form of eq. (\ref{eq:v}) except not containing the RCS. After reformulating the problem as the linear representation described above, any sparse reconstruction algorithm, such as LASSO, OMP, or FOCUSS, can be employed to estimate the target angles \cite{sparse}. In the multi-target scenario, this formulation can be applied  for each range parameter to extract the corresponding angular information. Notably, the computational complexity of this approach scales linearly with the number of targets, offering significant performance improvements in the asymptotic regime.
\section{Results}
    In this section, we present the studies that we conducted for the proposed system. In our simulations, we have considered a single-antenna radar operating at 240GHz with 10GHz bandwidth in a  coordinate system with metric axial dimensions. The radar is placed at $p_{\text{Tx}} = [0, 4]^T$. The metasurface pair modeled as 256  -element uniform linear arrays and (ULA) placed at $p_{\Psi_1} = [-0.1, 0]^T$, and $p_{\Psi_2} = [0.1, 0]^T$ respectively. When excited by the radar, the fixed phase shifts of the surfaces assumed to be providing maximum gains at $15^\circ$ and $75^\circ$ angles respectively. The simulations were performed at a signal-to-noise ratio (SNR) of 20dB at the receiver, while all the target RCS's randomly assigned between (0,1).

\begin{figure}[ht]
    \centering
    \includegraphics[width=0.80\linewidth]{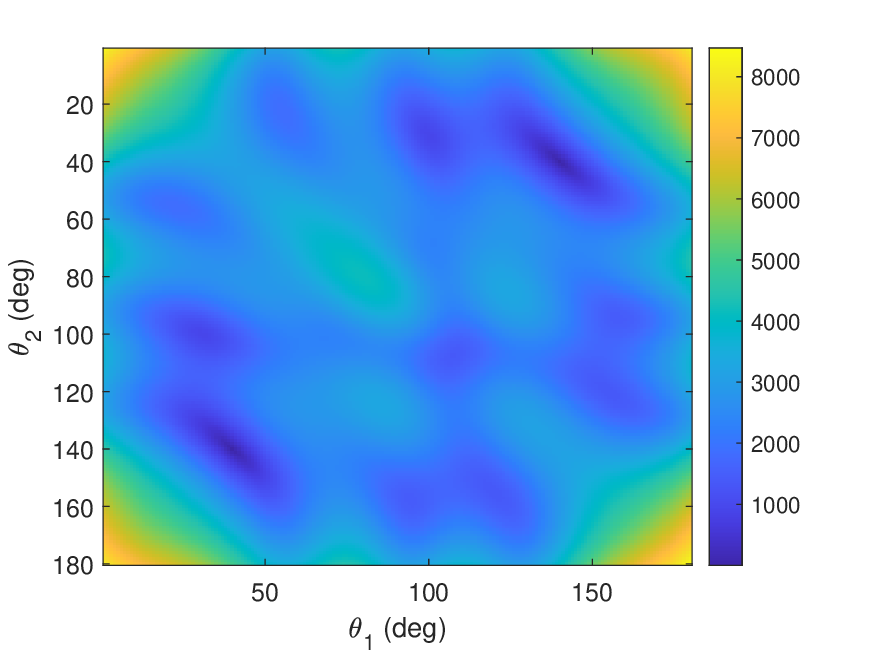}
    \caption{The loss function for multi-target simulation with the brute force search. The target headings are $\theta = \{40,140\}^\circ$.}
    \label{fig:bruteForce}
\end{figure}

To asses the performance with the brute-force search we simulated two targets with headings $\theta = [40, 140]^\circ$, placed $s_{1,m} = [2.41, 2.34]$ meters away from the metasurface pair. The resulting loss function from trying every configuration of  $(\theta_1,\theta_2)$ is given in the Fig. \ref{fig:bruteForce} the global minimum for the coordinates (40,140) is clearly observed. However, it is observed that the loss function is highly non-convex, which can lead to sub-optimal solutions, particularly in low-SNR conditions.

%The targets had the headings $\theta = [58,128,143]^\circ$, placed $s_n = [1.41, 2, 2.34]$ meters away from the preconfigured pair

\begin{figure}[ht]
    \centering
    \includegraphics[width=0.80\linewidth]{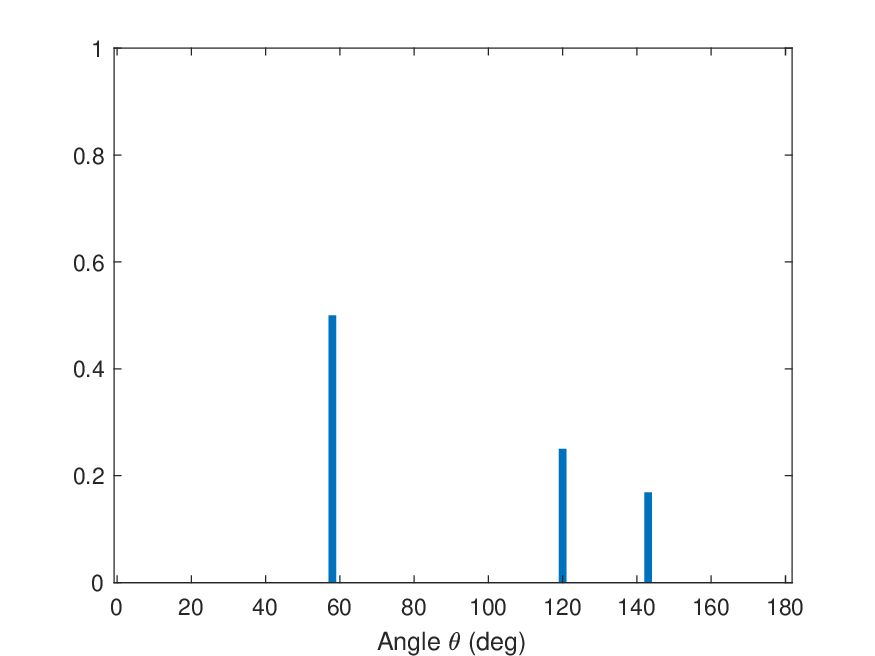}
    \caption{Estimated angles for multi-target simulation with LASSO method.} 
    \label{fig:lasso}
\end{figure}

Fig. \ref{fig:lasso} shows the estimated angles for the sparse reconstruction (LASSO) for three targets with the headings $\theta = [58, 128, 143]$ degrees, placed $s_{1,m} = [1.41, 2, 2.34]$ meters away from the metasurface pair. For each range parameter, a new sensing matrix is formed, and the method described in Section \ref{chp:cs} is run to extract a single support parameter. 

After demonstrating the effectiveness of the method, a discussion of the system parameters is needed. The sensing matrix constructed in sparse formulation provides valuable insights in this aspect. In compressed sensing theory, the coherence of a matrix is a measure of the largest correlation between any two columns of the matrix, defined as the maximum absolute value of the normalized inner product between distinct columns \cite{sparse}. It quantifies how similar the columns are to one another and plays a critical role in determining the matrix's suitability for sparse signal recovery. The parameters in (\ref{eq:v}) greatly effects the coherence properties of the sensing matrix. With the insight obtained from compressed sensing theory, we expect the columns of the sensing matrix to be as different as possible, to get a matrix with low coherence. In Figure 5 the sensing matrix for 4-element pair is presented across $10$GHz. We observe the sensing matrix can provide a unique representation between angles $5^\circ$ to $150^\circ$. With the disadvantage of low gain, which would greatly reduce the  operating range of the system. The signal power returning from the pair could be improved by increasing the number of elements, as can be seen in the Figure 6, where the 256-element pair is plotted. We observe a drastic improvement on the values of the sensing matrix for the beamforming angles $(\Psi_1=15^\circ, \Psi_2 = 75^\circ)$. However, this limits the operating directivity, as only a certain region around these angles would provide sufficient power. Moreover, unique recovery is now a problem with this sensing matrix, as most columns look-alike. Thanks to broad bandwiths available in the THz region, the system can be operated across a broad range of frequencies. The sensing matrix of the 256-element pair across 60 GHz is presented in Figure 7. In this case, due to beam squint effect, the pattern produced by the fixed phase shifts move with the frequency, breaking the coherence among the columns in the sensing matrix, making the sparse recovery effective again.

\begin{figure} [ht]
    \centering
    \includegraphics[width=0.80\linewidth]{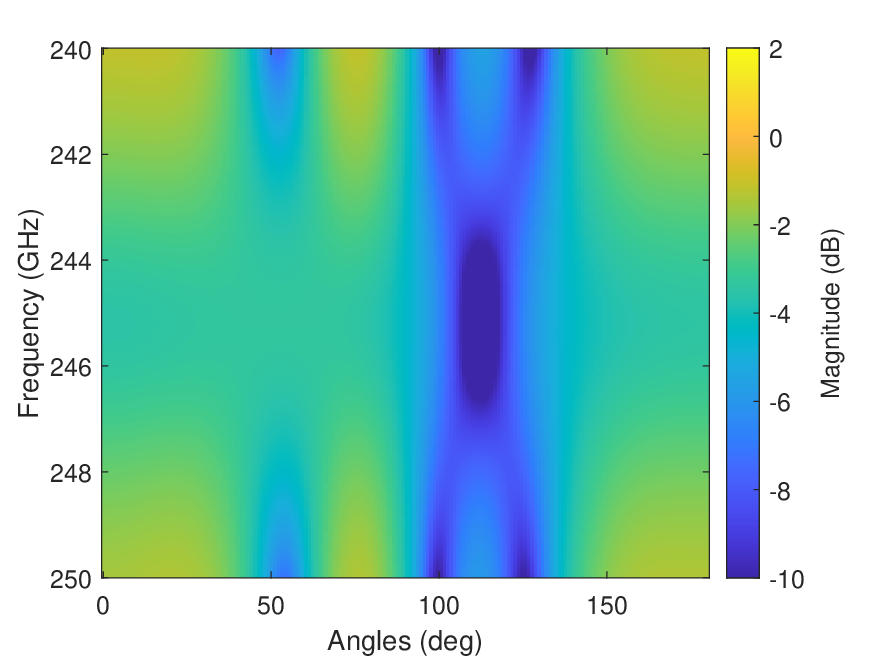}
    \caption{The sensing matrix for of the setup with 4 elements.}
    \includegraphics[width=0.80\linewidth]{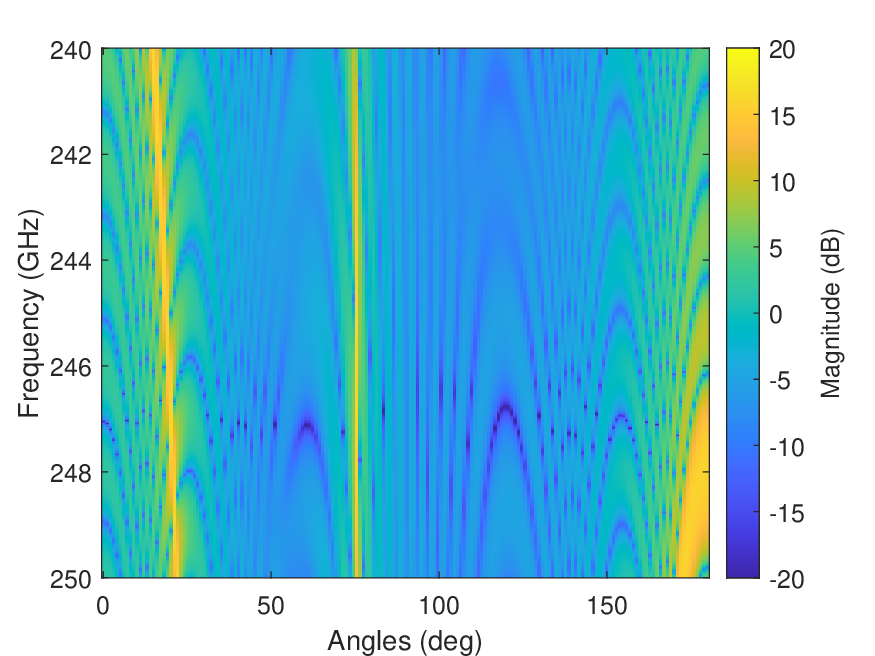}
    \caption{The sensing matrices for of the setup with 256 elements.}
    \includegraphics[width=0.80\linewidth]{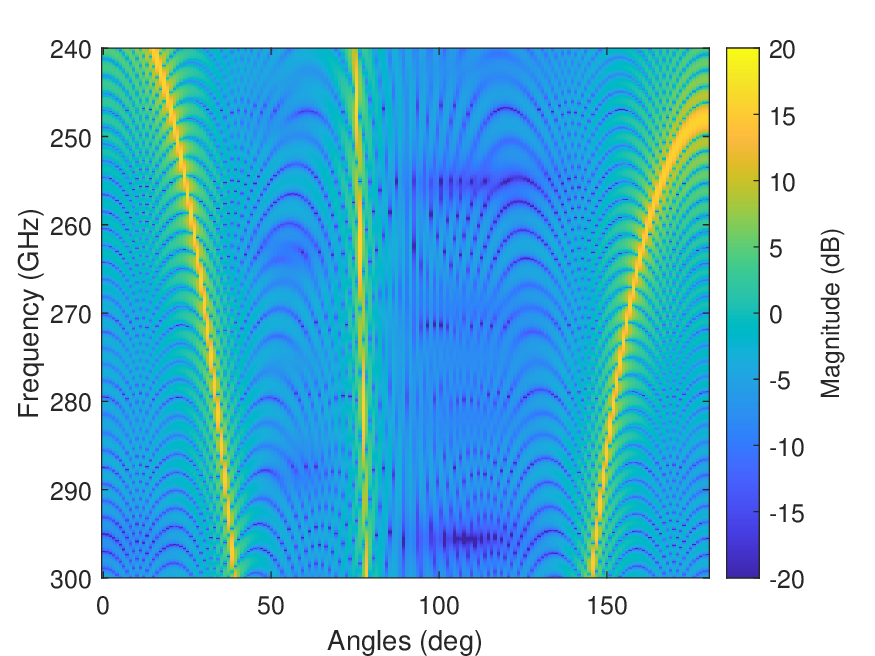}
    \caption{The sensing matrices for of the setup with 256 elements with 60 GHz bandwidth.}
    \label{fig:sensing_matrix}
\end{figure}

\section{Conclusions}
    In this study, we have presented a novel perspective for angular estimation in a single antenna radar utilizing the interference pattern produced by a pair of preconfigured metasurfaces. We have shown that after the range estimation, it is possible to extract an angular estimate by assessing the dependence between the target angle and the frequency on the interference pattern. The proposed method is capable of performing an angular estimation for multiple targets, achieving full localization with minimal infrastructure.

%The next natural steps in this research involve developing methods to design phase configurations that minimize the coherence of the sensing matrix while maximizing scene coverage. Furthermore, it is essential to investigate the global convergence properties of the sparse reconstruction method, in an effort to facilitate the construction of a unified sensing matrix that can effectively address the entire problem, accommodating varying target range parameters, $p_m$.

\bibliographystyle{IEEEtran}
\bibliography{references}

\end{document}